# Comprehensive Measurement of Three-Dimensional Thermal Conductivity Tensor Using a Beam-Offset Square-Pulsed Source (BO-SPS) Approach

Tao Chen[1], Shangzhi Song[1], Puqing Jiang[1],*

[1]*School of Power and Energy Engineering, Huazhong University of Science and Technology, Wuhan, Hubei 430074, China*

**ABSTRACT:** Accurately measuring the three-dimensional thermal conductivity tensor is essential for understanding and engineering the thermal behavior of anisotropic materials. Existing methods often struggle to isolate individual tensor elements, leading to large measurement uncertainties and time-consuming iterative fitting procedures. In this study, we introduce the Beam-Offset Square-Pulsed Source (BO-SPS) method for comprehensive measurements of three-dimensional anisotropic thermal conductivity tensors. This method uses square-pulsed heating and precise temperature rise measurements to achieve high signal-to-noise ratios, even with large beam offsets and low modulation frequencies, enabling the isolation of thermal conductivity tensor elements. We demonstrate and validate the BO-SPS method by measuring X-cut and AT-cut quartz samples. For X-cut quartz, with a known relationship between in-plane and cross-plane thermal conductivities, we can determine the full thermal conductivity tensor and heat capacity simultaneously. For AT-cut quartz, assuming a known heat capacity, we can determine the entire anisotropic thermal conductivity tensor, even with finite off-diagonal terms. Our results yield consistent principal thermal conductivity values for both quartz types, demonstrating the method's reliability and accuracy. This research highlights the BO-SPS method's potential to advance the understanding of thermal behavior in complex materials.

## I. INTRODUCTION

Accurate measurement of the anisotropic thermal conductivity tensor holds significant importance across various scientific and engineering disciplines. For instance, in electronics and semiconductor industries, where heat dissipation is critical for device reliability and efficiency, precise knowledge of thermal conductivity in different directions is indispensable for thermal management strategies [1]. Similarly, in materials science and engineering, anisotropic thermal conductivity plays a critical role in determining the heat transfer behavior of composite materials, layered structures, and crystalline solids [2, 3]. Moreover, accurate measurement of the anisotropic thermal conductivity tensor enables researchers to advance a fundamental understanding of heat conduction mechanisms in complex materials, thereby paving the way for developing innovative materials with tailored thermal properties for diverse applications ranging from energy storage and conversion to aerospace engineering [4, 5]. However, accurately measuring the anisotropic thermal conductivity of materials remains a challenge due to the complex nature of heat transfer in materials with directional dependencies.

In recent years, optical pump-probe techniques have become popular for measuring anisotropic thermal conductivities due to their high flexibility and ease of operation. Various optical methods have emerged, including beam-offset time-domain thermoreflectance (BO-TDTR) [6, 7], beam-offset frequency-domain thermoreflectance (BO-FDTR) [8, 9], and spatial-domain thermoreflectance (SDTR) [10]. These methods enable the measurement of arbitrarily aligned in-plane thermal conductivity tensors using offset pump and probe beams but have limitations. BO-TDTR and BO-FDTR can measure in-plane thermal conductivities only above $10\ \text{W}/(\text{m}\cdot\text{K})$ [8, 11], while SDTR extends the range down to $1\ \text{W}/(\text{m}\cdot\text{K})$ but cannot measure cross-plane thermal conductivity [10]. Additionally, these methods cannot completely isolate the in-plane thermal conductivity tensor elements, thus requiring time-consuming iterative fitting processes.

Isolating an in-plane thermal conductivity tensor element typically requires a large beam offset distance, exceeding five times the laser spot radius, and a low modulation frequency of around 100 Hz. These challenging conditions compromise the acquisition of high signal-to-noise ratio (SNR) data in conventional optical techniques. Alternatively, an advanced 3ω method employing multiple intricately designed heater lines meets these conditions and can measure all six elements of the thermal conductivity tensor [12]. However, this method requires a sufficiently large sample to accommodate heater lines and lacks flexibility.

In this study, we introduce the Beam-Offset Square-Pulsed Source (BO-SPS), an innovative all-optical method designed to overcome the limitations of existing techniques. By utilizing square-pulsed heating and precise acquisition of temperature rise amplitudes, the BO-SPS method achieves high SNR measurements even at large beam offset distances (exceeding five times the laser spot radius) and a significantly low modulation frequency of 100 Hz, enabling the complete isolation of thermal conductivity tensor elements.

The paper is structured as follows: First, we detail the experimental setup and measurement procedures for the BO-SPS method. This is followed by a deeper exploration, including the mathematical model, sensitivity analysis, and uncertainty estimation. Finally, we demonstrate the efficacy of this approach by accurately determining the thermal conductivity tensors of X-cut and AT-cut quartz samples.

## II. METHODOLOGIES

### 2.1 Experimental setup and measurement procedures

The operational concept of the SPS method has been previously established [13]. Essentially, SPS uses a square-wave-modulated pump beam to periodically heat the sample surface, while a probe beam measures the resulting temperature rise amplitude over time. The normalized temperature rise amplitude, plotted against normalized data acquisition time, is then fitted to a theoretical thermal model to extract the unknown thermal parameters.

A schematic diagram of our SPS system is shown in Fig. 1(a). The pump laser (Coherent OBIS LX FP 458 nm) and the probe laser (Thorlabs S4FC785) have wavelengths of 458 nm and 785 nm, respectively. The pump laser is electrically modulated using a square-wave function with a 50% duty cycle at a frequency of $f_0$ via a function generator. The modulated pump beam is directed by a dichroic mirror and then focused on the sample surface through an objective lens. The samples are coated with a thin metal transducer layer, usually ~100 nm thick aluminum (Al) film, to absorb the pump heat and provide a large thermoreflectance coefficient at the probe wavelength. The probe beam, passing through the same dichroic mirror, is focused on the sample surface. Precision control of the offset distance between the pump and probe is achieved by a pair of high-resolution motorized actuators tilting the dichroic mirror. The reflected probe beam is captured by a photodiode detector, and its output is processed by a periodic waveform analyzer (PWA), an advanced component of the UHF lock-in amplifier from Zurich Instruments. This setup efficiently yields the temperature rise amplitude with a high SNR over one heating period.

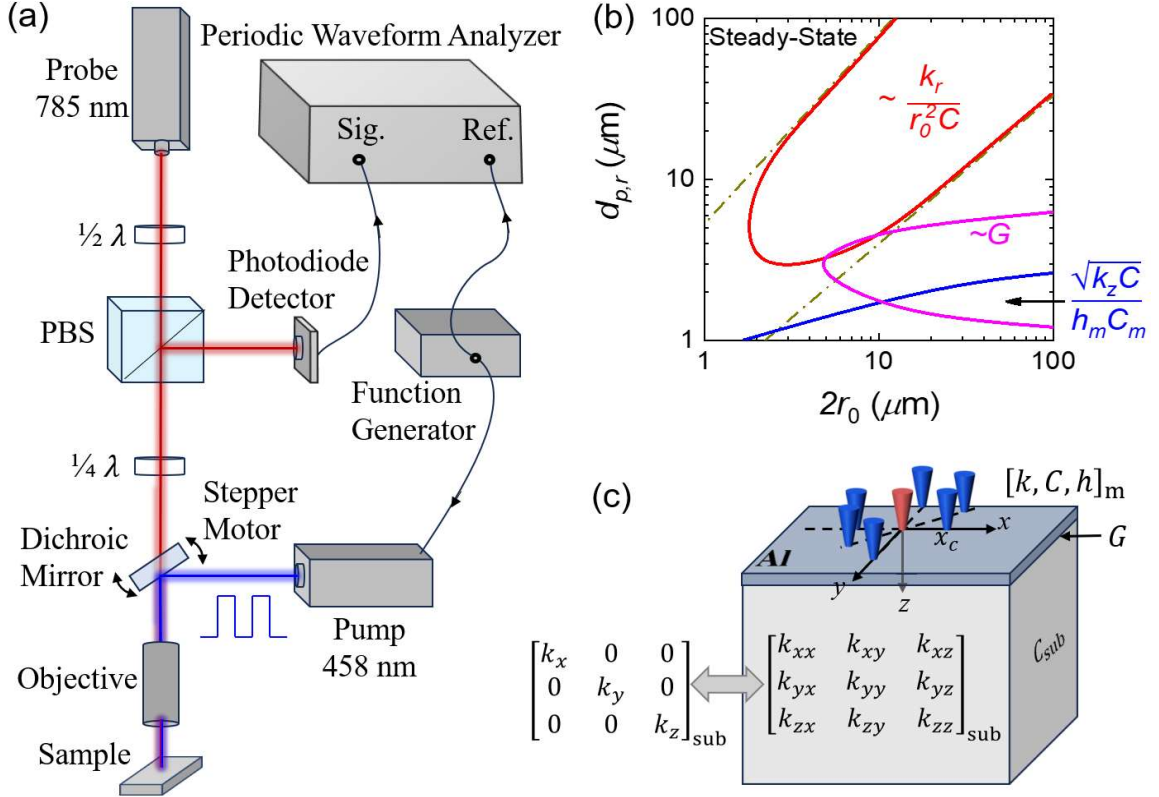

**FIG. 1.** (a) Schematic diagram of the experimental setup for the BO-SPS method. (b) Heat conduction mode with overlapped pump and probe spots, determined by two length scales: the in-plane thermal diffusion length $d_{p,r}$ and the laser spot diameter $2r_0$. (c) Illustration of the beam offset schemes used to determine the full thermal conductivity tensor of an anisotropic material.

Accurately determining the full thermal conductivity tensor of an anisotropic sample involves several steps, each requiring the optimal selection of specific parameters including the laser spot size $r_0$, the square-wave modulation frequency $f_0$, and the pump-probe offset distance $x_c$. Here, $r_0$ represents the root-mean-squared average of the $1/e^2$ radii of the pump and probe spots. These selections are crucial for ensuring measurement accuracy at each stage.

The first step is to measure the averaged in-plane thermal diffusivity $k_r/C$ and the cross-plane thermal effusivity $\sqrt{k_{zz}C}$ with overlapped pump and probe spots. The averaged in-plane thermal conductivity $k_r$ is related to the tensor elements as $k_r = \sqrt{k_{xx}k_{yy} - k_{xy}^2}$, and $C$ represents the volumetric heat capacity of the sample. Figure 1(b) shows the conduction mode when the pump and probe spots are overlapped. This mode is primarily influenced by comparing two length scales: the in-plane diffusion length $d_{p,r}$, defined as $d_{f,r} = \sqrt{k_r/\pi f_0 C}$, and the laser spot diameter $2r_0$. With a carefully selected modulation frequency and spot size such that $d_{p,r} \approx 2r_0$, the measured signal is mainly sensitive to $k_r/Cr_0^2$. Conversely, with a large spot size and a high modulation frequency such that $d_{f,r} \ll r_0/3$, heat transfer is primarily one-dimensional along the cross-plane direction, making the signal mainly sensitive to $\frac{\sqrt{k_{zz}C}}{h_m C_m}$. Here, $h_m$ and $C_m$ denote the thickness and volumetric heat capacity of the metal transducer film, respectively. Intermediate high-frequency measurements are also sensitive to the interfacial thermal conductance $G$ between the metal transducer film and the sample, allowing $G$ to be determined through multiple high-frequency measurements.

The next step involves determining different tensor elements through measurements with offset pump and probe spots in various directions. For a simple case with zero off-diagonal tensor elements, only one offset along the x-axis direction is sufficient to determine $k_{xx}$ and $k_{yy}$. In the most complex scenario, where all six tensor elements are non-zero and unknown, six different offset directions are necessary. These directions can be chosen as 0°, 180°, 90°, 270°, 45°, and 225°, as illustrated in Fig. 1(c). Repeated measurements in opposite offset directions are helpful to determine the off-diagonal elements. For example, if $k_{xz}$ is non-zero, offset measurements in the $+x$ and $-x$ directions will be sensitive to both $k_{xx}$ and $k_{xz}$ but in different manners: signals from the two counterpart measurements will exhibit the same sensitivity to $k_{xx}$ but opposite sensitivity to $k_{xz}$. Therefore, by fitting the product of the two sets of signals, $k_{xx}$ can be determined independently, while fitting their ratio allows for the independent determination of $k_{xz}$. Similarly, $k_{yz}$ can be determined by offset measurements in the $+y$ and $-y$ directions, and $k_{xy}$ can be determined by offset measurements in the 45° and 225° directions.

The offset distance $r_c = \sqrt{x_c^2 + y_c^2}$ and modulation frequency also need to be carefully selected to minimize sensitivity to tensor elements in the direction perpendicular to the offset, thereby allowing for the isolation of specific tensor elements. For most materials, using a low modulation frequency of $f_0 = 100$ Hz, a moderate laser spot size with $r_0 \leq 10\,\mu$m, and an offset distance of $r_c \approx 5r_0$ is sufficient to achieve this goal while maintaining a high SNR. These settings generally works well for materials with $k_r$ ranging from 1 to $1000$ W/(m·K). For materials with a lower $k_r$ less than $1$ W/(m·K), it is recommended to reduce the modulation frequency to around $10$ Hz and extend the offset distance to $r_c \approx 10r_0$. Additionally, to maintain a high sensitivity to $k_r$ of the substrate, it is advisable to select a metal transducer layer with a thermal conductivity less than 100 times the $k_r$ of the substrate. Conversely, for highly conductive materials with $k_r$ above $1000$ W/(m·K), the same offset distance of $r_c \approx 5r_0$ can be used, but the modulation frequency needs to be increased to $f_0 = 1$ kHz to achieve the same suppression of sensitivity to $k_{yy}$. By adhering to these guidelines, one should be able to effectively isolate the tensor elements and achieve accurate measurements across a wide range of material conductivities.

The measurement procedures described above will be demonstrated through exemplary measurements of X-cut and AT-cut quartz samples, detailed in Section III.

## 2.2 Three-dimensional thermal model

In a Cartesian three-dimensional orthogonal coordinate system, the thermal conductivity tensor **k** of a material can be represented as a $3 \times 3$ matrix in the form:

$$\mathbf{k} = \begin{bmatrix} k_{xx} & k_{xy} & k_{xz} \\ k_{yx} & k_{yy} & k_{yz} \\ k_{zx} & k_{zy} & k_{zz} \end{bmatrix} \tag{1}$$

In the absence of magnetic fields, the thermal conductivity tensor is symmetric, which means

$k_{xy} = k_{yx}$, $k_{xz} = k_{zx}$, and $k_{yz} = k_{zy}$. The Spectral Theorem states that this real symmetric matrix can be diagonalized by an orthogonal transformation as $\mathbf{k}' = \begin{bmatrix} k_x & 0 & 0 \\ 0 & k_y & 0 \\ 0 & 0 & k_z \end{bmatrix}$, which represents the thermal conductivity tensor in the coordinate system aligned with its principal axes.

Given this symmetry, the three-dimensional heat diffusion equation for a material with the thermal conductivity tensor $\mathbf{k}$ can be expressed as:

$$C \frac{\partial T}{\partial t} = k_{xx} \frac{\partial^2 T}{\partial x^2} + k_{yy} \frac{\partial^2 T}{\partial y^2} + k_{zz} \frac{\partial^2 T}{\partial z^2} + 2k_{xy} \frac{\partial^2 T}{\partial x \partial y} + 2k_{xz} \frac{\partial^2 T}{\partial x \partial z} + 2k_{yz} \frac{\partial^2 T}{\partial y \partial z} \quad (2)$$

where $C$ represents the volumetric heat capacity of the material, and $T$ is the temperature, which is a function of the spatial coordinates $x, y, z,$ and time $t$.

In SPS experiments, the pump beam has a Gaussian profile and is modulated by a square wave, applying a surface heat flux given by:

$$p_0(x, y, t) = \frac{2A_1}{\pi \sigma_{x_1} \sigma_{y_1}} \exp\left(-\left(\frac{2x^2}{\sigma_{x_1}^2} + \frac{2y^2}{\sigma_{y_1}^2}\right)\right) \left(\frac{1}{2} + \frac{2}{\pi} \sum_{n=1}^{\infty} \frac{\sin(2\pi(2n-1)f_0 t)}{2n-1}\right) \quad (3)$$

where $A_1$ is the average power of the pump beam absorbed by the sample surface; $\sigma_{x_1}$ and $\sigma_{y_1}$ are the $1/e^2$ radii of the pump spot in the $x$ and $y$ directions, respectively.

The Fourier transform of $p_0(x, y, t)$ over space and time is:

$$P_0(u, v, \omega) = A_1 \exp\left(-\frac{\pi^2 u^2 \sigma_{x_1}^2}{2}\right) \exp\left(-\frac{\pi^2 v^2 \sigma_{y_1}^2}{2}\right) \left(\frac{\delta(\omega)}{2} + \frac{1}{\pi} \sum_{n=1}^{\infty} i \frac{(\delta(\omega + 2\pi(2n-1)f_0) - \delta(\omega - 2\pi(2n-1)f_0))}{2n-1}\right) \quad (4)$$

Here, $\delta(\omega)$ is the Dirac delta function, $u$ and $v$ are the spatial frequency variables, and $\omega$ is the temporal frequency variable.

Another continuous-wave laser beam, with an offset distance of $(x_c, y_c)$ relative to the pump beam, is used to detect the area-weighted average of the transient temperature changes. The detected temperature variation over time can be expressed as:

$$\Delta T(t) = \int_{-\infty}^{\infty} \int_{-\infty}^{\infty} \int_{-\infty}^{\infty} \hat{G}(u, v, \omega) P_0(u, v, \omega) e^{i\omega t} d\omega \exp\left(-\frac{\pi^2 u^2 \sigma_{x_2}^2}{2}\right) \exp\left(-\frac{\pi^2 v^2 \sigma_{y_2}^2}{2}\right) e^{i2\pi(ux_c + vy_c)} du dv \quad (5)$$

where, $\sigma_{x_2}$ and $\sigma_{y_2}$ are the $1/e^2$ radii of the probe spot in the x and y directions, respectively; $x_c$ and $y_c$ are the offset distances of the probe beam relative to the pump beam in the x and y directions, respectively; $\hat{G}(u, v, \omega)$ is the Green's function for the temperature response, the full derivation of which can be found in Supplementary Information Section S1 and some previous literature [14, 15].

The numerical solution of $\Delta T(t)$, after further normalization in the same manner as the experimental data, can be employed for best fitting of the experimental data.

## 2.3 Sensitivity analysis

Sensitivity analysis is used to assess the impact of different parameters on the signal in a system. The sensitivity coefficient is defined as:

$$S_\xi = \frac{\partial \ln A_{\text{norm}}}{\partial \ln \xi} \tag{6}$$

where $\xi$ represents any parameter to be analyzed, and $A_{\text{norm}}$ is the normalized amplitude signal. If the magnitude of the sensitivity coefficient is greater than 0.2, the signal is considered highly sensitive to that parameter. Conversely, if the magnitude of the sensitivity coefficient is less than 0.05, the sensitivity is considered low, indicating that the signal is minimally affected by variations in that parameter.

Sensitivity analysis is essential both before and after experiments. Before experiments, it guides the optimal selection of the pump modulation frequency $f_0$, laser spot size $r_0$, and offset distance $(x_c, y_c)$ to minimize the measurement uncertainty. After experiments, it helps to estimate the measurement uncertainty.

## 2.4 Uncertainty analysis

This study employs the least squares method to fit experimental data, and when simultaneously fitting $M$ groups of signals, the squared loss function $J$ is defined as follows:

$$J = \prod_{j=1}^{M} \sqrt{\frac{1}{N_j}\sum_{i=1}^{N_j}\left(\frac{g_j(X_U,X_P,t_i)}{y_j(t_i)}-1\right)^2} \tag{7}$$

Here, $j$ indicates the $j$-th set of experimental signals, $y_j(t_i)$ represents the normalized amplitude signal measured at the $i$-th time point $t_i$ for the $j$-th set, $N_j$ is the total number of data points for the $j$-th set. The function $g_j(X_U,X_P,t_i)$ represents the predicted value at time $t_i$ based on the thermal model with predictors $X_U$ and $X_P$ for the $j$-th set, where $X_U$ is the vector of unknown parameters and $X_P$ is the vector of known parameters.

During the optimal fitting process, the unknown parameter vector $\hat{X}_U$ is determined by minimizing the squared loss function such that the partial derivative of the loss function to each unknown parameter $u_l$ equals zero:

$$\frac{\partial J(X_U,X_P)}{\partial u_l}\Big|_{\hat{X}_U} = 0, \quad \text{for } l = 1,2,\dots,\text{length}(X_U) \tag{8}$$

Solving this equation yields the covariance matrix of $\hat{X}_U$ as:

$$\text{Var}[\hat{X}_U] = \begin{bmatrix} \sigma_{u_1}^2 & \text{cov}[u_1,u_2] & \cdots \\ \text{cov}[u_2,u_1] & \sigma_{u_2}^2 & \cdots \\ \vdots & \vdots & \ddots \end{bmatrix} \tag{9}$$

Here, the elements on the main diagonal represent the variance of each unknown parameter. A $2\sigma$ interval (corresponding to a 95% confidence level) is used to assess the uncertainty of the unknown parameters. The matrix adequately accounts for experimental noise and the uncertainty of the input parameters. Further details can be found in reference [16] and Supplementary Information Section S2.

In this work, the input parameters include the thickness $h_m$, thermal conductivity $k_m$, and heat capacity $C_m$ of the metal transducer film, the spot size $r_0$, and offset distance $(x_c, y_c)$. Among these parameters, $h_m$ can be determined using a step profiler with an uncertainty of 5%; $k_m$ can be derived from the electrical resistivity measured using a four-point probe method, with an uncertainty of 5%; $C_m$ can be obtained from literature databases, with a typical uncertainty of less than 3%; $r_0$ can be determined using the knife-edge method with an uncertainty of 2%; the calibration of $x_c$ and $y_c$ can be performed using resolution calibration plates, providing precise reference points for spatial accuracy. For enhanced reliability, these parameters are also calibrated by measuring a standard fused silica sample alongside the unknown sample using the same BO-SPS method. The standard silica sample was placed side-by-side with the unknown sample during the metal film deposition process and coated with the same metal film. This approach ensures consistent calibration and minimizes measurement uncertainty.

The heat capacity $C$ of the sample under test can be determined concurrently with its thermal conductivity if any of its in-plane thermal conductivities is known to be equal to its cross-plane thermal conductivity. However, if this equivalence is not known, $C$ must be treated as a known input parameter. Typically, $C$ can be sourced from the literature database, with an uncertainty of approximately 3%.

The final source of uncertainty is noise in the signals. However, through meticulous experimental design, we can attain measurements with exceptionally high SNR, even with a large beam offset distance exceeding five times the spot size at an ultra-low modulation frequency of 100 Hz. This rigorous design ensures that the uncertainty due to noise remains below 1%.

## III. RESULTS AND DISCUSSION

The efficacy of the BO-SPS method is demonstrated through measurements of X-cut and AT-cut quartz crystal samples, encompassing four scenarios with varying levels of complexity: 1) X-cut quartz with its *c*-axis aligned parallel to the *x*-coordinate axis; 2) X-cut quartz with its *c*-axis oriented in an arbitrary in-plane direction; 3) AT-cut quartz with its *c*-axis positioned in the *xz* plane; and 4) AT-cut quartz with its *c*-axis not aligned with any specific coordinate plane. The sample orientations and coordinate systems are illustrated in Fig. 2. These samples, which are of optical grade, were purchased from MTI. All the samples are coated with a nominal 100 nm thick Al transducer layer for thermoreflectance measurements.

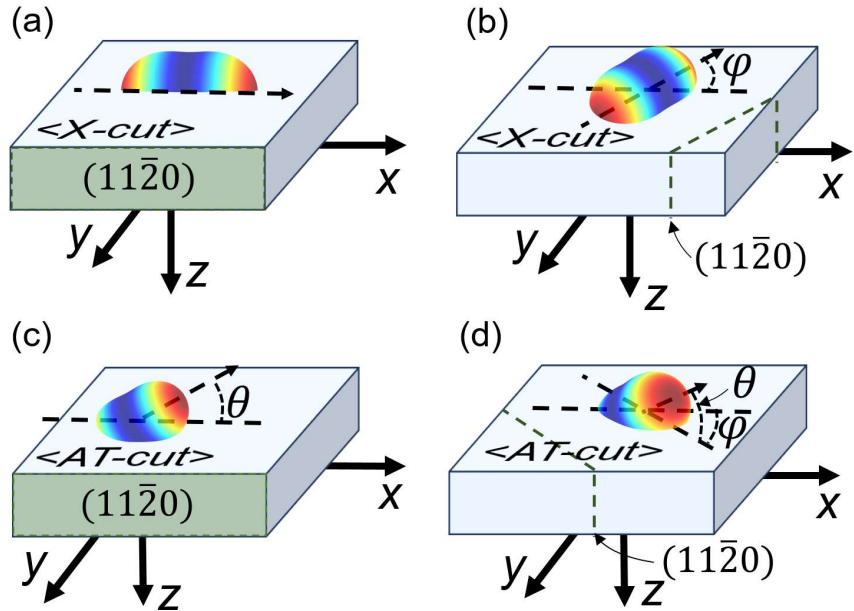

**Fig. 2** Illustration of the orientations of quartz samples measured in this study with varying complexity: (a) X-cut quartz with its *c*-axis aligned parallel to the *x*-coordinate axis; (b) X-cut quartz with its *c*-axis oriented at an angle $\varphi$ with the *x*-coordinate axis; (c) AT-cut quartz with its *c*-axis positioned in the *xz* plane; and (d) AT-cut quartz with its *c*-axis not aligned with any specific coordinate plane. The $(11\bar{2}0)$ plane, which runs parallel to the *c*-axis, is also depicted in the images for additional insight into the orientation of the quartz samples.

### 3.1 Case 1: X-cut quartz with *c*-axis aligned parallel to *x*-coordinate axis

For the X-cut quartz sample, with its *c*-axis aligned parallel to the *x*-coordinate axis, the thermal conductivity tensor has specific relationships: $k_{xx} = k_c$, $k_{yy} = k_{zz} = k_a$, and $k_{xy} = k_{xz} = k_{yz} = 0$. Therefore, there are only two unknown elements in the tensor: $k_{xx}$ and $k_{yy}$.

To determine the thermal conductivity tensor of this sample, we first use a laser spot size of $r_0 =$

8.3 $\mu$m, a modulation frequency of $f_0 = 100$ Hz, and an offset distance of $x_c = 40$ $\mu$m in the x-direction for the measurement. The measured signals for one full period, along with the best-fit thermal model predictions, are shown in Fig. 3(a1). A magnified portion of the data is displayed in Fig. 3(a2), covering the normalized time range of 0.01-0.1 on a log-log scale, facilitating a clearer view of the signal fitting quality. Model predictions with $\pm 30\%$ bounds of the best-fit $k_{xx}$ value demonstrate the high sensitivity of this signal to $k_{xx}$ of the sample. A comprehensive sensitivity analysis, as depicted in Fig. 3(a3), indicates that this set of signals is primarily sensitive to the combined parameter $\frac{k_{xx}}{Cx_c^2}$. This relationship is derived because the sensitivity coefficient for $k_{xx}$ has the same magnitude but opposite sign as that for $C$, and the sensitivity coefficient for $x_c$ is twice that for $C$. With $x_c$ carefully calibrated before the experiment, the best fit for this set of signals yields $k_{xx}/C$ of the sample, with a value of $0.0557 \pm 0.0008$ cm$^2$/s.

Subsequently, we repeat the measurement using the same laser spot size of $r_0 = 8.3$ $\mu$m, a higher modulation frequency of $f_0 = 1$ kHz, and zero offset distances ($x_c = y_c = 0$) for the measurement. The measured signals and sensitivity coefficients are presented in Fig. 3(b1-b3). Sensitivity analysis suggests that this set of signals is predominantly sensitive to the combined parameter $\frac{\sqrt{k_{xx}k_{yy}}}{Cr_0^2}$, as evidenced by the sensitivity coefficients for these parameters, indicating $S_{k_{xx}} = S_{k_{yy}} = -0.5S_C = -0.25S_{r_0}$. With $r_0$ carefully calibrated before the experiment, the best fit for this set of signals yields $\sqrt{k_{xx}k_{yy}}/C$ of the sample, with a value of $0.0429 \pm 0.0015$ cm$^2$/s.

The third set of signals, as illustrated in Fig. 3(c1-c3), involves using a larger laser spot size of $r_0 = 18$ $\mu$m, a significantly high modulation frequency of $f_0 = 9$ MHz, and zero offset distances for the measurement. Sensitivity coefficients plotted in Fig. 3(c3) suggest that this set of signals is predominantly sensitive to the combined parameter $\frac{\sqrt{k_{zz}C}}{h_m C_m}$, with some additional sensitivity to the combined parameter $h_m/k_{z,m}$, since the following relationships among the sensitivity coefficients for these parameters are manifested: $S_{k_{zz}} = S_C = -0.5 S_{C_m}$, and $S_{C_m} - S_{h_m} = S_{k_{z,m}}$. With the thermal conductivity $k_{z,m}$, heat capacity $C_m$, and thickness $h_m$ of the metal transducer layer carefully calibrated before the experiment, the best fit for this set of signals yields $\sqrt{k_{zz} C}$ of the sample, with a value of $0.3524 \pm 0.0060 \text{ J}/(\text{s}^{0.5} \cdot \text{cm}^2 \cdot \text{K})$.

Up to this point, the above three sets of measurements provide $k_{xx}/C$, $\sqrt{k_{xx} k_{yy}}/C$, and $\sqrt{k_{zz} C}$ for the sample, respectively. Combined with the prior knowledge that $k_{xx} = k_c$ and $k_{yy} = k_{zz} = k_a$, we can determine the properties of the X-cut quartz sample as $k_c = 10.8 \pm 0.48 \text{ W}/(\text{m} \cdot \text{K})$, $k_a = 6.4 \pm 0.25 \text{ W}/(\text{m} \cdot \text{K})$, and $C = 1.94 \pm 0.08 \text{ MJ}/(\text{m}^3 \cdot \text{K})$.

In the literature [10, 17-20], reported values for $k_c$ range from 10 to 11.8 $\text{W}/(\text{m} \cdot \text{K})$, and values for $k_a$ range from 6 to 7 $\text{W}/(\text{m} \cdot \text{K})$. The Thermophysical Properties Research Center (TPRC) database [21] recommends $k_c = 10.4 \text{ W}/(\text{m} \cdot \text{K})$ and $k_a = 6.21 \text{ W}/(\text{m} \cdot \text{K})$ for high-quality quartz single crystals at 300 K, with an uncertainty of 5%. Variations in reported values for the heat capacity of quartz crystals are relatively small. Anderson [22] reported a specific heat of $0.736 \text{ J}/(\text{g} \cdot \text{K})$ at 296 K and a density of $2.6378 \text{ g}/\text{cm}^3$, converting to a volumetric heat capacity of $1.9414 \text{ J}/(\text{cm}^3 \cdot \text{K})$. Our current measurements align well with these accepted values.

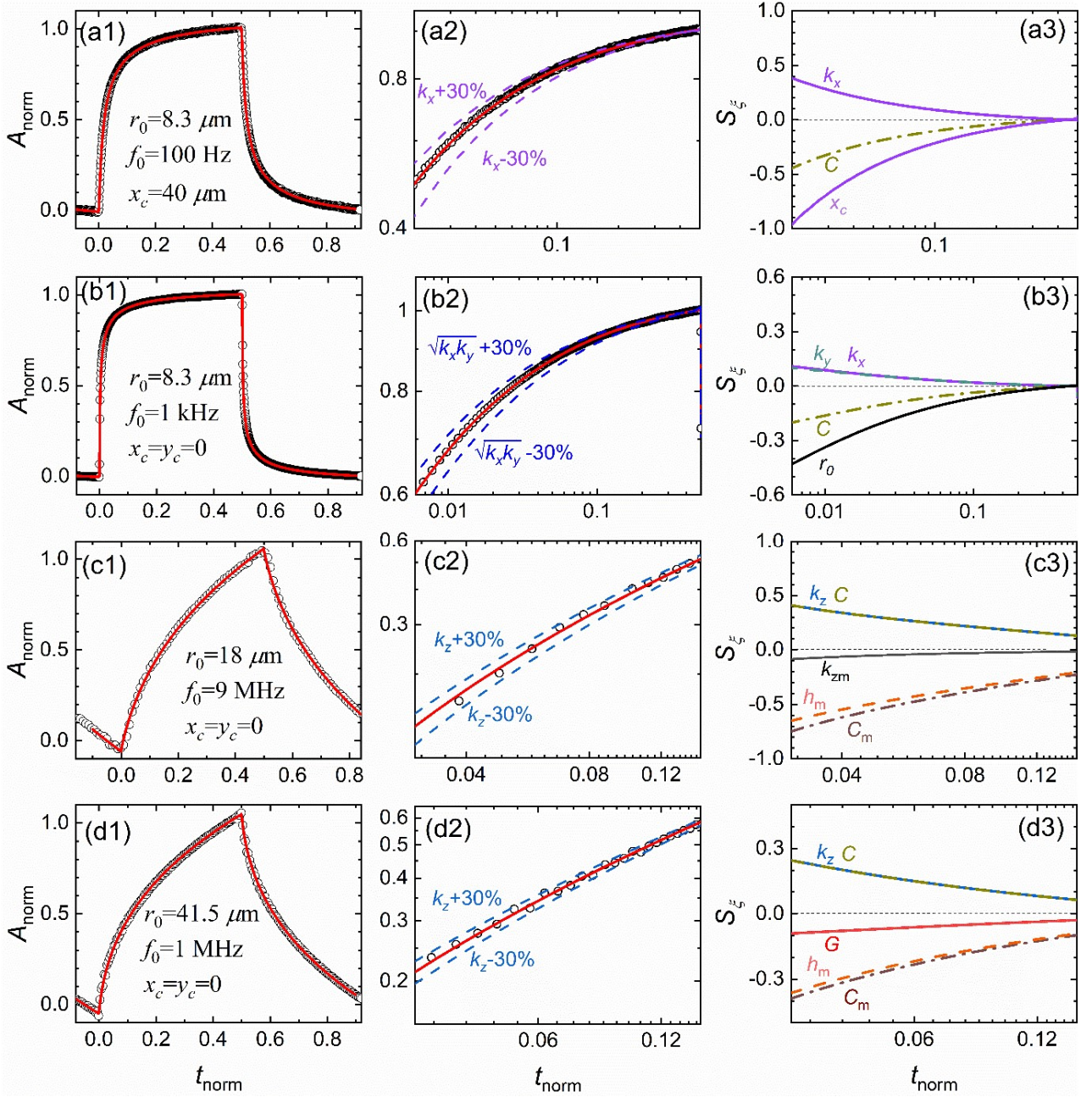

**FIG. 3.** Four sets of measurements to determine the full thermal conductivity tensor and heat capacity of an X-cut quartz sample with its *c*-axis aligned parallel to the *x*-axis. Specifically, (a1-d1) shows the measured signals over one period alongside the best-fitted model predictions, (a2-d2) presents a zoomed-in portion of the signals in a log-log scale to demonstrate the fitting quality, and (a3-d3) shows the sensitivity coefficients of the signals in (a2-d2) to the parameters in the thermal model. Measurement configurations for each set are as follows: (a1-a3): $r_0 = 8.3\ \mu m$, $f_0 = 100$ Hz, $x_c = 40\ \mu m$; (b1-b3): $r_0 = 8.3\ \mu m$, $f_0 = 1$ kHz, $x_c = y_c = 0\ \mu m$; (c1-c3): $r_0 = 18\ \mu m$, $f_0 = 9$ MHz, $x_c = y_c = 0\ \mu m$; (d1-d3): $r_0 = 41.5\ \mu m$, $f_0 = 1$ MHz, $x_c = y_c = 0\ \mu m$.

The signals from the three sets of measurements mentioned above are not sensitive to the thermal conductance $G$ of the Al/quartz interface. Therefore, determining $G$ is not necessary for calculating the thermal conductivity tensor and heat capacity of the quartz sample. However, if $G$ is desired, it can be determined from a fourth set of measurements, which involves employing a significantly large laser spot size of $r_0 = 41.5\ \mu$m, a moderately high modulation frequency of $f_0 = 1$ MHz, and zero offset distances. The measured signals and sensitivity coefficients are plotted in Fig. 3(d1-d3). Sensitivity analysis suggests that this set of signals is sensitive to both the combined parameter $\frac{\sqrt{k_{zz}C}}{h_m C_m}$ and $h_m C_m/G$, given the relationships among the sensitivity coefficients for these parameters: $S_{k_{zz}} = S_C$, $S_{h_m} = S_{C_m}$, and $2S_{k_z} + S_G = -S_{h_m}$. With $\sqrt{k_{zz}C}$ and $h_m C_m$ pre-determined, the best fit for this set of signals yields $G$ for the sample, which is $G = 110 \pm 20$ MW/(m² · K) for the Al/quartz interface.

### 3.2 Case 2: X-cut quartz with c-axis in arbitrary in-plane orientation

Next, we explore a more generalized scenario in which the c-axis of X-cut quartz is randomly oriented in the xy-plane. For verification purposes, the sample was intentionally rotated counterclockwise by a known angle of 30° compared to Case 1. In this configuration, the principal axis of the X-cut quartz sample is no longer aligned with any of the coordinate axes, resulting in a non-zero off-diagonal component $k_{xy}$ for the in-plane thermal conductivity tensor.

To determine the thermal conductivity tensor of this sample, we first use a laser spot size of $r_0 = 8.7\ \mu$m, a modulation frequency of $f_0 = 5$ kHz, and zero offset distances for the measurement, as illustrated in Fig. 4(a1-a3), where both the measured signals and sensitivity coefficients are depicted. Given the absence of any offset in this particular set of measurements, the coordinate system can also be defined with the x-axis parallel to the c-axis of X-cut quartz with no influence on the outcomes.

Within this coordinated framework, similar to the previous measurements outlined in Fig. 3(b), this collection of signals primarily reflects sensitivity toward the combined parameter $\frac{\sqrt{k_x k_y}}{C r_0^2}$, where $k_x$ and $k_y$ denote the principal thermal conductivity tensor elements in the new coordinate system and are interconnected with the tensor elements in the current coordinate system through $\sqrt{k_x k_y} = \sqrt{k_{xx} k_{yy} - k_{xy}^2}$. With $r_0$ carefully calibrated before the experiment, the optimal fitting of the signal set in Fig. 4(a2) provides the $\sqrt{k_x k_y}/C$ value of the sample, which is $0.0429 \pm 0.0015$ cm²/s.

Using the same laser spot size, a low modulation frequency of 100 Hz, and offsetting the pump and probe spots in the $x$-direction for 40 $\mu$m, the signals become dominantly sensitive to the combined parameter $\frac{k_{xx}}{C x_c^2}$, as shown in Fig. 4(b1-b3). With $x_c$ carefully calibrated before the experiment, the best fit for this set of signals yields $k_{xx}/C$ of the sample, with a value of $0.0500 \pm 0.0005$ cm²/s.

Using the same laser spot size and modulation frequency, but offsetting the pump and probe spots in the $y$-direction for 40 $\mu$m, the signals become dominantly sensitive to the combined parameter $\frac{k_{yy}}{C y_c^2}$, as shown in Fig. 4(c1-c3). With $y_c$ carefully calibrated before the experiment, the best fit for this set of signals yields $k_{yy}/C$ of the sample, with a value of $0.0387 \pm 0.0005$ cm²/s.

Up to this point, the three sets of measurements mentioned above yield $\sqrt{k_x k_y}/C$, $k_{xx}/C$, and $k_{yy}/C$ for the sample, respectively. Consequently, the off-diagonal term of in-plane thermal diffusivity can be derived as $\frac{k_{xy}}{C} = \pm \sqrt{\frac{k_{xx}}{C} \frac{k_{yy}}{C} - \frac{k_x k_y}{C^2}}$, where the sign requires further verification. To determine the sign of the $k_{xy}/C$ term, we conduct another set of measurements using the same laser spot size and modulation frequency but offsetting the pump and probe spots in the 45° direction for 40 $\mu$m. The measured signals and sensitivity coefficients are shown in Fig. 4(d1-d3). We observe that the measured signals can only be fitted when $k_{xy}/C$ is negative, while the model predictions assuming a positive $k_{xy}/C$ deviate significantly from the measured signals, as illustrated in Fig. 4(d2).

Therefore, $k_{xy}/C$ of this sample is determined as $-0.0098 \pm 0.0001$ cm$^2$/s.

The principal in-plane thermal diffusivities $k_x/C$ and $k_y/C$ of the sample can be obtained through an orthogonal transformation of the in-plane thermal diffusivity tensor as:

$$\begin{bmatrix} \cos\varphi & -\sin\varphi \\ \sin\varphi & \cos\varphi \end{bmatrix}^T \begin{bmatrix} k_{xx}/C & k_{xy}/C \\ k_{xy}/C & k_{yy}/C \end{bmatrix} \begin{bmatrix} \cos\varphi & -\sin\varphi \\ \sin\varphi & \cos\varphi \end{bmatrix} = \begin{bmatrix} k_x/C & 0 \\ 0 & k_y/C \end{bmatrix} \quad (10)$$

Here, $\varphi$ is the angle of rotation in the $xy$ plane from the current coordinate system to the new coordinate system and can be expressed explicitly as: $\varphi = \frac{1}{2}\arctan\left(2k_{xy}/(k_{xx} - k_{yy})\right)$. From the measured in-plane thermal diffusivity tensor, we calculate the rotation angle as $\varphi = -29.97° \pm 0.39°$, which matches closely with the nominal value of $-30°$. Diagonalization yields the principal in-plane thermal diffusivities as:

$$\begin{bmatrix} 0.8663 & 0.4995 \\ -0.4995 & 0.8663 \end{bmatrix}^T \begin{bmatrix} 0.0500 & -0.0098 \\ -0.0098 & 0.0387 \end{bmatrix} \begin{bmatrix} 0.8663 & 0.4995 \\ -0.4995 & 0.8663 \end{bmatrix} = \begin{bmatrix} 0.0556 & 0 \\ 0 & 0.0330 \end{bmatrix} \text{ (cm}^2\text{/s)} \quad (11)$$

With the principal in-plane thermal diffusivities determined, combining the through-pane thermal effusivity from the measurement shown in Fig. 3(c) and the prior knowledge that $k_x = k_c$ and $k_y = k_{zz} = k_a$, the principal elements of the thermal conductivity tensor and the heat capacity can thus be derived as $\mathbf{k}' = \begin{bmatrix} 10.8 & 0 & 0 \\ 0 & 6.4 & 0 \\ 0 & 0 & 6.4 \end{bmatrix}$ W/(m·K), along with the heat capacity determined as $C = 1.94$ MJ/(m$^3$·K). These values are consistent with the results from Case 1 measurements.

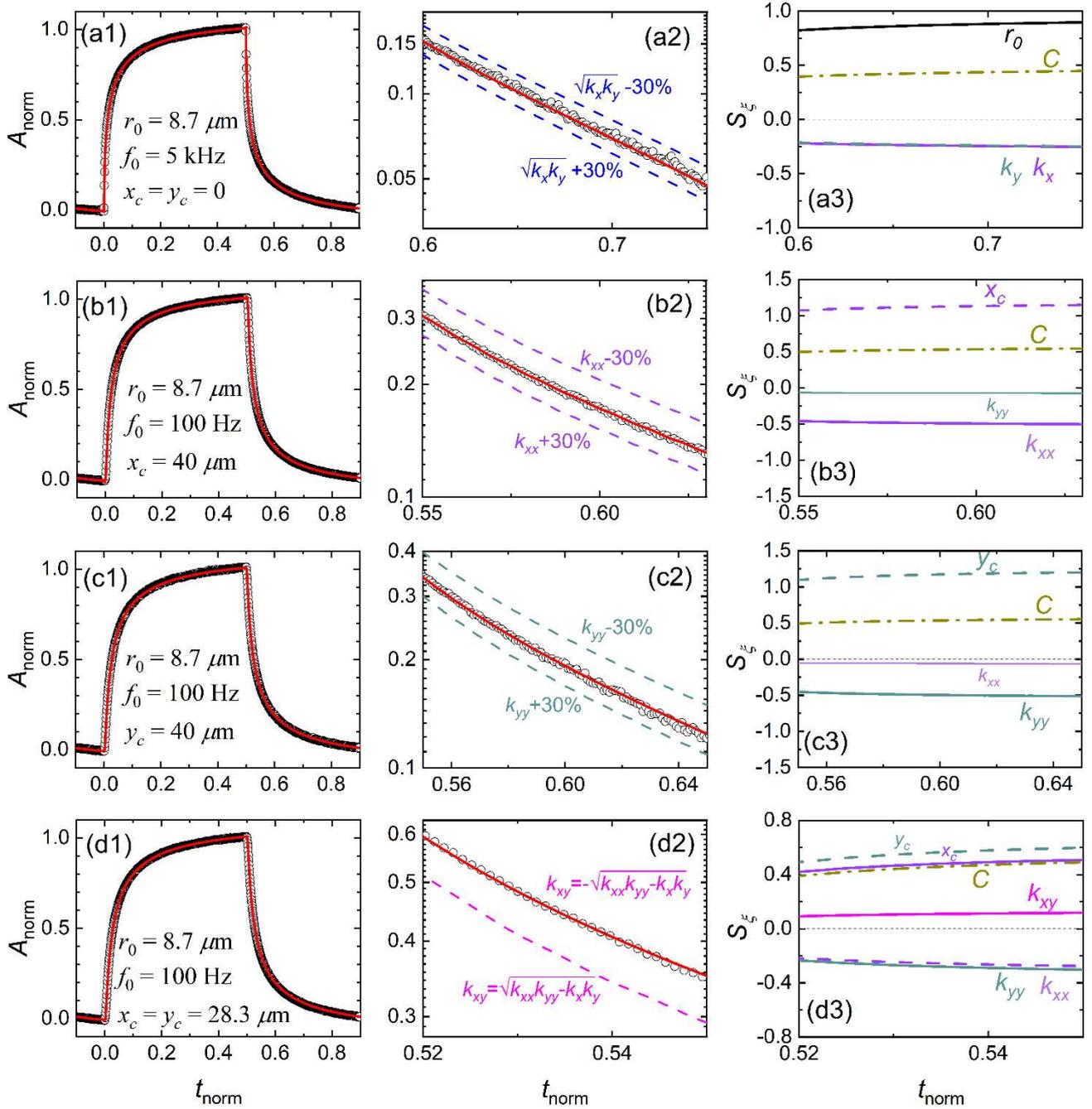

**Fig. 4.** Four sets of measurements to determine the in-plane thermal conductivity tensor of an X-cut quartz sample with its *c*-axis in an arbitrary in-plane direction. Specifically, (a1-d1) shows the measured signals over one period alongside the best-fitted model predictions, (a2-d2) presents a zoomed-in portion of the signals in a log-log scale to demonstrate the fitting quality, and (a3-d3) shows the sensitivity coefficients of the signals in (a2-d2) to the parameters in the thermal model. Measurement configurations for each set are as follows: (a1-a3): $r_0 = 8.7$ $\mu$m, $f_0 = 5$ kHz, $x_c = y_c = 0$ $\mu$m; (b1-b3): $r_0 = 8.7$ $\mu$m, $f_0 = 100$ Hz, $x_c = 40$ $\mu$m; (c1-c3): $r_0 = 8.7$ $\mu$m, $f_0 = 100$ Hz, $x_c = 40$ $\mu$m; (d1-d3): $r_0 = 8.7$ $\mu$m, $f_0 = 100$ Hz, $x_c = y_c = 28.3$ $\mu$m.

### 3.3 Case 3: AT-cut quartz with *c*-axis in *xz*-plane

For an AT-cut quartz crystal, the cut is made at an angle of 35.25° relative to the c-axis in the a-axis plane. Therefore, for the AT-cut quartz sample measured in this study, the c-axis of the sample has an inclined angle of 35.25° with the xy-plane in our coordinate system. When the sample is aligned such that its c-axis is in the xz-plane, only $k_{xz}$ is nonzero while the other two, $k_{xy}$ and $k_{yz}$, are zero.

To determine the thermal conductivity tensor of this sample, we first use a laser spot size of $r_0 = 8.0\ \mu m$, a modulation frequency of $f_0 = 100$ Hz, and an offset distance of $y_c = 40\ \mu m$ in the y-direction for the measurement, with the measured signals and sensitivity coefficients depicted in Fig. 5(a1-a3). This set of signals is predominantly sensitive to the combined parameter $\frac{k_{yy}}{Cy_c^2}$, as suggested by the sensitivity coefficients in Fig. 5(a3). With $y_c$ carefully calibrated before the experiment and $C$ treated as an input parameter, the best fit for this set of signals yields $k_{yy}$ of the sample, which is $6.40 \pm 0.24$ W/(m·K).

The second set of measurements involves using the same laser spot size, a slightly higher modulation frequency of 2 kHz, and zero offset distances. This collection of signals is dominantly sensitive to $\frac{\sqrt{k_x k_y}}{Cr_0^2}$, as shown in Fig. 5(b1-b3). With $r_0$ carefully calibrated before the experiment and $C$ treated as an input parameter, the optimal fitting of this signal set yields $\sqrt{k_x k_y} = 7.71 \pm 0.31$ W/(m·K) for this sample. We note that $\sqrt{k_x k_y} = \sqrt{k_{xx} k_{yy}}$ since the off-diagonal term $k_{xy}$ is zero for this sample. Therefore, with $k_{yy}$ predetermined from the last set of measurements, we can now determine $k_{xx}$ as $9.3 \pm 0.9$ W/(m·K).

In the third and fourth sets of measurements, we use the same laser spot size of 8.0 μm and a low modulation frequency of 100 Hz, offsetting the pump and probe spots in the positive and negative x-directions by 40 μm respectively, with the measured signals and sensitivity coefficients depicted in Fig. 5(c1-c3) and (d1-d3). These signals are sensitive not only to $\frac{k_{xx}}{Cx_c^2}$ but also to $k_{xz}$, as depicted in Fig. 5(c3) and (d3). With pre-determined $k_{xx}$ from the first two sets of measurements and $C$ as an input parameter, the best fit for these two sets of signals yields $k_{xz} = -2.05 \pm 0.09 \text{ W/(m·K)}$ for this sample. Since these two sets of signals have the same sensitivity to $k_{xx}$ but opposite sensitivity to $k_{xz}$, both $k_{xx}$ and $k_{xz}$ can be independently determined by simultaneously fitting these two sets of signals. The $k_{xx}$ values independently determined from measurements in Fig. 5(c, d) and Fig. (a, b) agree with each other, reinforcing the accuracy of the current measurements.

The last step involves determining $\sqrt{k_{zz}C}$, which can be achieved by measurements using the same spot size, a high modulation frequency of 9 MHz, and zero offset distances, with the measured signals and sensitivity coefficients depicted in Fig. 5(e1-e3). This set of signals is dominantly sensitive to $\frac{\sqrt{k_{zz}C}}{h_m C_m}$, as shown in Fig. 5(e3). With the metal film's areal heat capacitance $h_m C_m$ carefully calibrated and $C$ as an input parameter, the best fit for this set of signals yields $k_{zz} = 7.87 \pm 0.55 \text{ W/(m·K)}$ for this sample.

With all the tensor elements determined, an orthogonal transformation of the full thermal diffusivity tensor yields:

$$\begin{bmatrix} \cos\theta & 0 & -\sin\theta \\ 0 & 1 & 0 \\ \sin\theta & 0 & \cos\theta \end{bmatrix}^T \begin{bmatrix} 9.3 & 0 & -2.05 \\ 0 & 6.4 & 0 \\ -2.05 & 0 & 7.87 \end{bmatrix} \begin{bmatrix} \cos\theta & 0 & -\sin\theta \\ 0 & 1 & 0 \\ \sin\theta & 0 & \cos\theta \end{bmatrix} = \begin{bmatrix} 10.7561 & 0 & 0 \\ 0 & 6.4 & 0 \\ 0 & 0 & 6.4139 \end{bmatrix} \text{(W/(m·K))} \quad (12)$$

where $\theta$ is the rotation angle in the xz-plane and can be determined as $\theta = \frac{1}{2}\arctan(2k_{xz}/(k_{xx} - k_{zz}))$. Therefore, from the comprehensive measurements of the full thermal conductivity tensor in this case, we determine the principal thermal conductivities for AT-cut quartz as $k_c = 10.76 \pm 0.41 \text{ W/(m·K)}$, $k_a = 6.4 \pm 0.31 \text{ W/(m·K)}$, which are consistent with the values of the X-cut quartz, and the angle $\theta$ as $\theta = -35.39° \pm 2.38°$, which matches closely with the theoretical value of $-35.25°$.

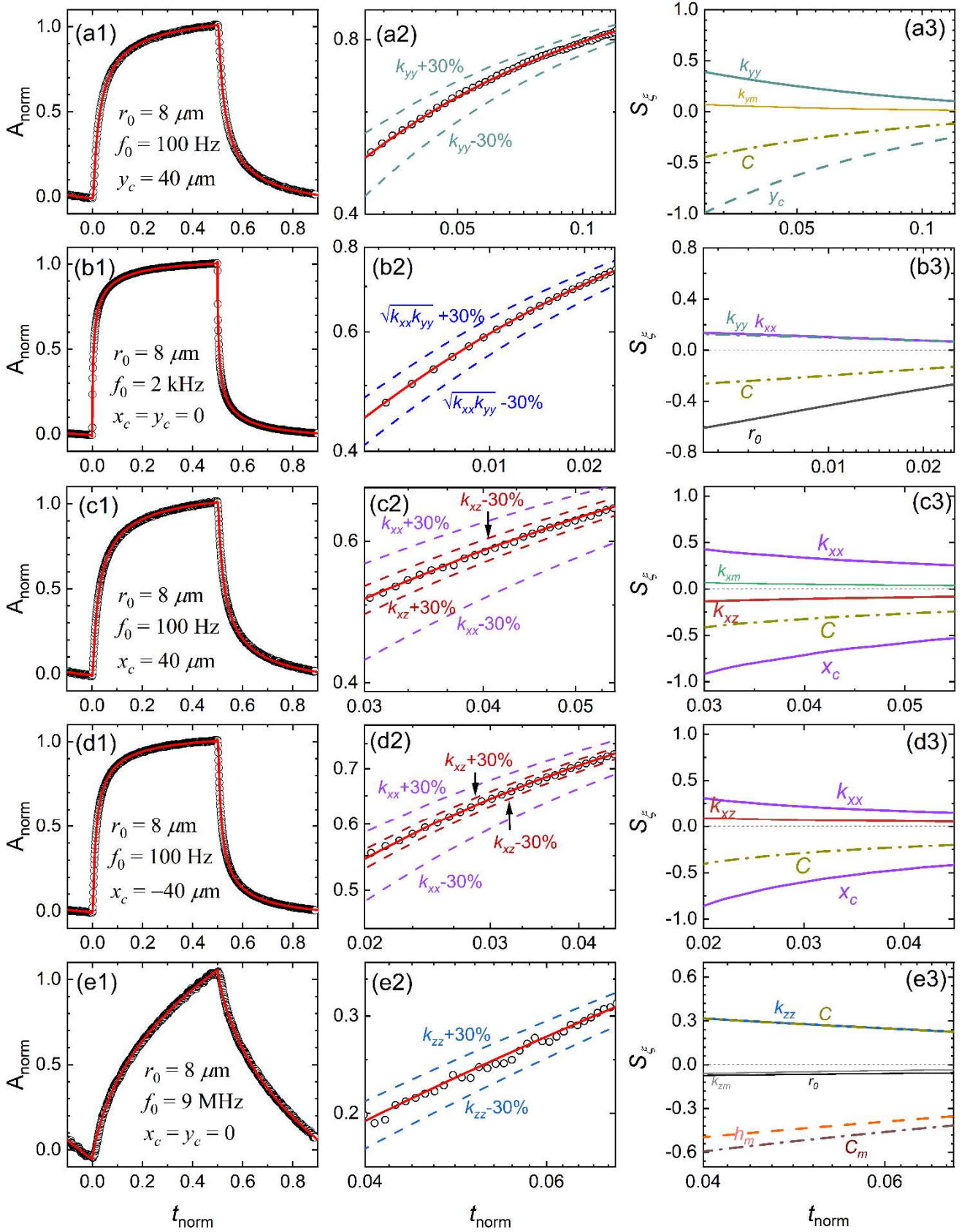

**Fig. 5**. Five sets of measurements to determine the full thermal conductivity tensor of an AT-cut quartz sample with its *c*-axis aligned in the *xz*-plane. Specifically, (a1-e1) shows the measured signals over one period alongside the best-fitted model predictions, (a2-e2) presents a zoomed-in portion of the signals in a log-log scale to demonstrate the fitting quality, and (a3-e3) shows the sensitivity coefficients of the signals in (a2-e2) to the parameters in the thermal model. Measurement configurations for each set are as follows: (a1-a3): $r_0 = 8.0\ \mu m$, $f_0 = 100$ Hz, $y_c = 40\ \mu m$; (b1-b3): $r_0 = 8.0\ \mu m$, $f_0 = 2$ kHz, $x_c = y_c = 0\ \mu m$; (c1-c3): $r_0 = 8.0\ \mu m$, $f_0 = 100$ Hz, $x_c = 40\ \mu m$; (d1-d3): $r_0 = 8.0\ \mu m$, $f_0 = 100$ Hz, $x_c = -40\ \mu m$; (e1-e3): $r_0 = 8.0\ \mu m$, $f_0 = 9$ MHz, $x_c = y_c = 0\ \mu m$.

### 3.4 Case 4: AT-cut quartz with *c*-axis in arbitrary 3D orientation

Lastly, we tackle the most complex scenario involving AT-cut quartz with its *c*-axis not aligned with any specific coordinate plane. For verification purposes, we deliberately rotate the AT-cut quartz sample clockwise by 45° about the *z*-axis. All six elements of the thermal conductivity tensor are nonzero in this case and need to be determined.

To achieve this goal, we first use a laser spot size of $r_0 = 8.0\ \mu m$, a modulation frequency of $f_0 = 2$ kHz, and zero offset distances. This collection of signals is dominantly sensitive to $\frac{\sqrt{k_x k_y}}{C r_0^2}$, as shown in Fig. 6(a1-a3). With $r_0$ carefully calibrated and $C$ treated as an input parameter, the optimal fitting of this signal set yields $\sqrt{k_x k_y} = 7.71 \pm 0.31$ W/(m · K) for this sample.

In the second and third sets of measurements, we use the same laser spot size of $r_0 = 8.0\ \mu m$ and a low modulation frequency of $f_0 = 100$ Hz, offsetting the pump and probe spots by 40 $\mu$m in the positive and negative *x*-directions, respectively, with the measured signals and sensitivity coefficients displayed in Fig. 6(b1-b3) and (c1-c3). These signals are sensitive not only to $\frac{k_{xx}}{C x_c^2}$ but also to $k_{xz}$. However, the sensitivity coefficients for $k_{xz}$ have opposite signs for these two counterpart measurements, while the sensitivity coefficients for $k_{xx}$ remain the same. Therefore, both $k_{xx}$ and $k_{xz}$ can be determined through the best fit of these two sets of signals, which are $k_{xx} = 7.85 \pm 0.29$ W/(m · K) and $k_{xz} = -1.45 \pm 0.07$ W/(m · K) for this sample.

In the fourth and fifth sets of measurements, we use the same laser spot size of $r_0 = 8.0\ \mu m$ and a low modulation frequency of $f_0 = 100$ Hz, offsetting the pump and probe spots by 45 $\mu$m in the positive and negative *y*-directions, respectively, with the measured signals and sensitivity coefficients displayed in Fig. 6(d1-d3) and (e1-e3). These signals are sensitive not only to $\frac{k_{yy}}{Cy_c^2}$ but also to $k_{yz}$. Similarly, because the sensitivity coefficients for $k_{yz}$ have opposite signs, different from those for $k_{yy}$ for these two counterpart measurements, the best fit for these two sets of signals yields $k_{yy} = 7.85 \pm 0.29 \text{ W/(m·K)}$ and $k_{yz} = -1.45 \pm 0.07 \text{ W/(m·K)}$, respectively, for this sample.

Up to this point, the sample's in-plane tensor elements $k_x k_y$, $k_{xx}$, and $k_{yy}$ have been determined. Consequently, the in-plane off-diagonal term can be derived as $k_{xy} = \pm\sqrt{k_{xx}k_{yy} - k_x k_y}$, where the sign requires further verification. To determine the sign of $k_{xy}$, we used the same laser spot size of $r_0 = 8.0 \text{ }\mu\text{m}$ and modulation frequency of $f_0 = 100 \text{ Hz}$ for the measurement, offsetting the pump and probe spots by 40 $\mu$m in the 45° and 225° directions, respectively. Each set of signals is sensitive to $k_{xy}$, $k_{xz}$, and $k_{yz}$ simultaneously. However, the sensitivity coefficients for $k_{xz}$ and $k_{yz}$ have opposite signs for the two counterpart measurements, while the sensitivity coefficients for $k_{xy}$ remain the same. Therefore, the product of these two sets of signals is no longer sensitive to $k_{xz}$ and $k_{yz}$ but still sensitive to $k_{xy}$. Figure 8(f1-f3) shows the product of these two counterpart measurements and the corresponding sensitivity coefficients. We observe that the measured signals can only be fitted when $k_{xy}$ is set as positive, while the model predictions assuming a negative $k_{xy}$ deviate significantly from the measured signals, as illustrated in Fig. 8(f2). Therefore, the $k_{xy}$ of the sample can be determined as $k_{xy} = 1.45 \pm 0.05 \text{ W/(m·K)}$.

The $k_{zz}$ measured in Fig. 5(e) from Case 3 still applies here, as rotating the sample about the *z*-axis does not affect the measurement of $k_{zz}$.

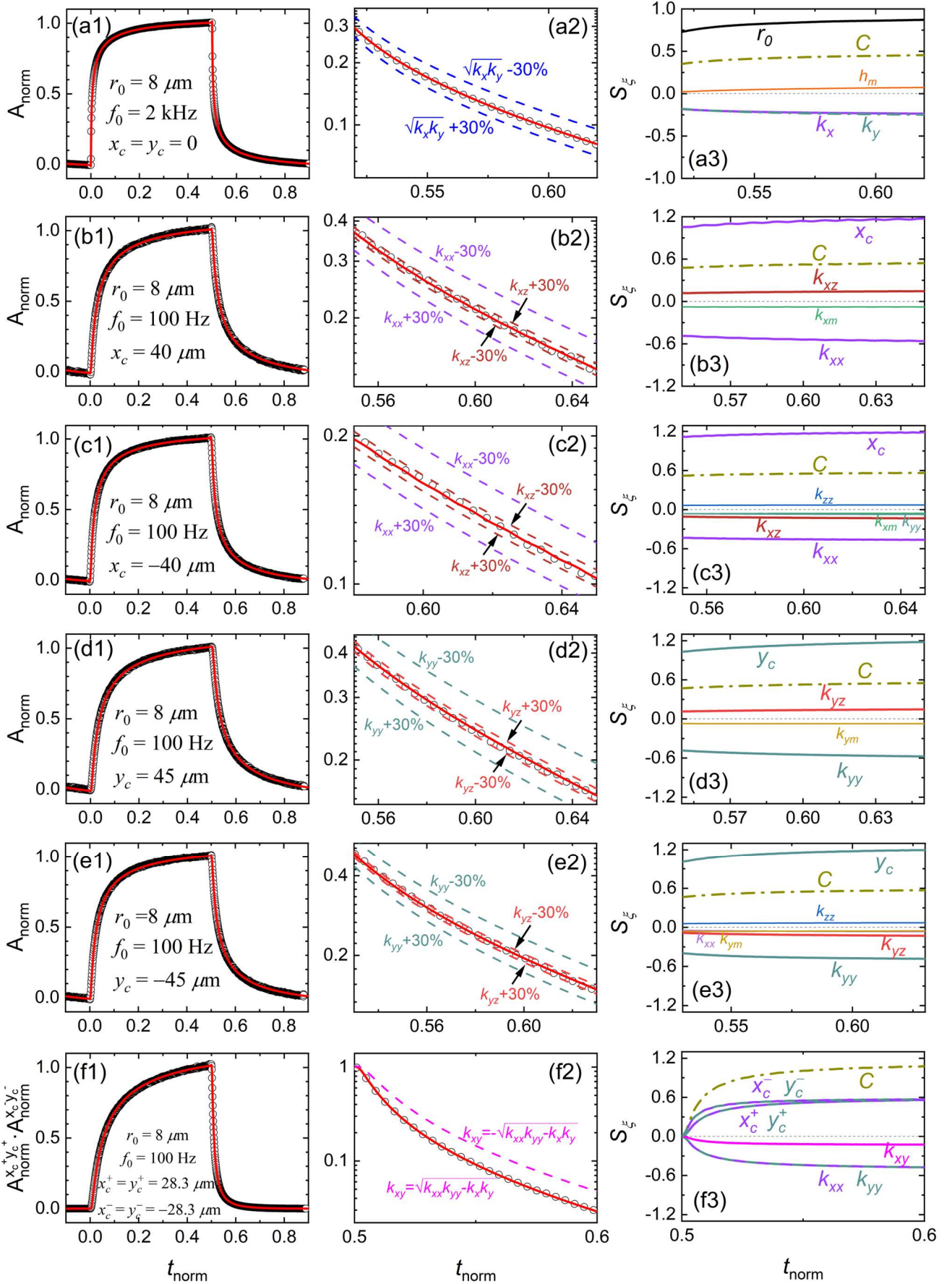

**Fig. 6**. Six sets of measurements to determine the full thermal conductivity tensor of an AT-cut quartz

sample with its *c*-axis not aligned with any coordinate plane. Specifically, (a1-f1) shows the measured signals over one period alongside the best-fitted model predictions, (a2-f2) presents a zoomed-in portion of the signals in a log-log scale to demonstrate the fitting quality, and (a3-f3) shows the sensitivity coefficients of the signals in (a2-f2) to the parameters in the thermal model. Measurement configurations for each set are as follows: (a1-a3): $r_0 = 8.0\ \mu m$, $f_0 = 2$ kHz, $x_c = y_c = 0\ \mu m$; (b1-b3): $r_0 = 8.0\ \mu m$, $f_0 = 100$ Hz, $x_c = 40\ \mu m$; (c1-c3): $r_0 = 8.0\ \mu m$, $f_0 = 100$ Hz, $y_c = -40\ \mu m$; (d1-d3): $r_0 = 8.0\ \mu m$, $f_0 = 100$ Hz, $y_c = 45\ \mu m$; (e1-e3): $r_0 = 8.0\ \mu m$, $f_0 = 100$ Hz, $y_c = -45\ \mu m$; (f1-f3): $r_0 = 8.0\ \mu m$, $f_0 = 100$ Hz, $x_c = y_c = \pm 28.3\ \mu m$.

With all the tensor elements determined, an orthogonal transformation of the full thermal diffusivity tensor yields:

$$\begin{bmatrix} \cos\theta & 0 & -\sin\theta \\ 0 & 1 & 0 \\ \sin\theta & 0 & \cos\theta \end{bmatrix}^T \begin{bmatrix} \cos\varphi & -\sin\varphi & 0 \\ \sin\varphi & \cos\varphi & 0 \\ 0 & 0 & 1 \end{bmatrix}^T \begin{bmatrix} 7.85 & 1.45 & -1.45 \\ 1.45 & 7.85 & -1.45 \\ -1.45 & -1.45 & 7.87 \end{bmatrix} \begin{bmatrix} \cos\varphi & -\sin\varphi & 0 \\ \sin\varphi & \cos\varphi & 0 \\ 0 & 0 & 1 \end{bmatrix} \begin{bmatrix} \cos\theta & 0 & -\sin\theta \\ 0 & 1 & 0 \\ \sin\theta & 0 & \cos\theta \end{bmatrix} =$$

$$\begin{bmatrix} 10.7567 & 0 & 0 \\ 0 & 6.4000 & 0 \\ 0 & 0 & 6.4133 \end{bmatrix} (W/(m\cdot K)) \quad (13)$$

Here, $\theta$ is the angle between the *c*-axis and the *xy*-plane, and $\varphi$ is the angle between the projection of the *c*-axis onto the *xy*-plane and the *x*-axis. From the comprehensive measurements of the full thermal conductivity tensor in this case, we determine the principal thermal conductivities of AT-cut quartz as $k_c = 10.76 \pm 0.40$ W/(m · K), $k_a = 6.4 \pm 0.31$ W/(m · K), and the angles as $\varphi = 45° \pm 1.45°$, $\theta = -35.4° \pm 2.34°$. Both the principal thermal conductivities and the angles align perfectly well with previous measurements.

## V. SUMMARY AND OUTLOOK

In this study, we have introduced and validated the Beam-Offset Square-Pulsed Source (BO-SPS) method for comprehensive measurement of three-dimensional anisotropic thermal conductivity tensors. By employing square-pulsed heating and precise temperature rise measurements, the BO-SPS method achieves high signal-to-noise ratios, even with large beam offsets and low modulation frequencies, allowing for complete isolation of thermal conductivity tensor elements. The application to X-cut and AT-cut quartz samples demonstrates the method's efficacy and reliability. For X-cut quartz, we successfully determined the full thermal conductivity tensor and heat capacity simultaneously, leveraging the known relationship between in-plane and cross-plane thermal conductivities. For AT-cut quartz, assuming a known heat capacity, we accurately measured the entire anisotropic thermal conductivity tensor, including finite off-diagonal terms. The consistency of the principal thermal conductivity values obtained for both quartz types further validates the accuracy of our method.

The BO-SPS method represents a major advancement in the study of anisotropic materials, effectively overcoming the limitations of conventional techniques. This breakthrough is essential for the development and optimization of materials in various technological applications. Future work will focus on extending the BO-SPS method to a wider range of anisotropic materials and exploring its potential in different temperature regimes and environmental conditions.

## ACKNOWLEDGMENTS

P.J. acknowledges the support of the National Natural Science Foundation of China (NSFC) through Grant No. 52376058.

## AUTHOR DECLARATIONS

### Conflict of Interest

The authors have no conflicts to disclose.

### Author Contributions

**Tao Chen:** Data curation (equal); Formal analysis (equal); Investigation (equal); Methodology (equal); Validation (equal); Writing – original draft (equal); Writing – review &editing (equal). **Shangzhi Song:** Investigation (equal); Methodology (equal). **Puqing Jiang:** Conceptualization (lead);

Formal analysis (equal); Methodology (equal); Validation (equal); Writing – original draft (equal); Writing – review & editing (equal).

DATA AVAILABILITY

The data that support the findings of this study are available from the corresponding authors upon reasonable request.

REFERENCES

[1] H. Chen, V.V. Ginzburg, J. Yang, Y. Yang, W. Liu, Y. Huang, L. Du, B. Chen, Thermal Conductivity of Polymer-Based Composites: Fundamentals and Applications, Prog. Polym. Sci., 59 (2016) 41-85.
[2] P. Xiao, E. Chávez-Ángel, S. Chaitoglou, M. Sledzinska, A. Dimoulas, C.M. Sotomayor Torres, A. El Sachat, Anisotropic Thermal Conductivity of Crystalline Layered SnSe2, Nano Lett., DOI (2021).
[3] H. Jang, C.R. Ryder, J.D. Wood, M.C. Hersam, D.G. Cahill, 3D Anisotropic Thermal Conductivity of Exfoliated Rhenium Disulfide, Adv. Mater., 29 (2017) 1700650.
[4] P. Ferrando-Villalba, A.F. Lopeandia, F.X. Alvarez, B.K. Paul, C.d. Tomas, M.I. Alonso, M. Garriga, A.R. Goñi, J. Santiso, G. Garcia, J. Rodríguez-Viejo, Tailoring thermal conductivity by engineering compositional gradients in Si1−xGex superlattices, Nano Research, 8 (2015) 2833-2841.
[5] H. Cao, S.-Z. Li, J. Yang, Z. Liu, L. Bai, W. Yang, Thermally Conductive Magnetic Composite Phase Change Materials for Anisotropic Photo/Magnetic-to-Thermal Energy Conversion, ACS applied materials & interfaces, DOI (2023).
[6] J.P. Feser, D.G. Cahill, Probing anisotropic heat transport using time-domain thermoreflectance with offset laser spots, Rev. Sci. Instrum., 83 (2012) 104901.
[7] J.P. Feser, J. Liu, D.G. Cahill, Pump-probe measurements of the thermal conductivity tensor for materials lacking in-plane symmetry, Rev. Sci. Instrum., 85 (2014) 104903.
[8] D. Rodin, S.K. Yee, Simultaneous measurement of in-plane and through-plane thermal conductivity using beam-offset frequency domain thermoreflectance, Rev. Sci. Instrum., 88 (2017) 014902.
[9] L. Tang, C. Dames, Anisotropic thermal conductivity tensor measurements using beam-offset frequency domain thermoreflectance (BO-FDTR) for materials lacking in-plane symmetry, Int. J. Heat Mass Transfer, 164 (2021) 120600.
[10] P. Jiang, D. Wang, Z. Xiang, R. Yang, H. Ban, A new spatial-domain thermoreflectance method to measure a broad range of anisotropic in-plane thermal conductivity, Int. J. Heat Mass Transfer, 191 (2022) 122849.
[11] P. Jiang, X. Qian, R. Yang, Time-domain thermoreflectance (TDTR) measurements of anisotropic thermal conductivity using a variable spot size approach, Rev. Sci. Instrum., 88 (2017) 074901.
[12] V. Mishra, C.L. Hardin, J.E. Garay, C. Dames, A 3 omega method to measure an arbitrary anisotropic thermal conductivity tensor, Rev. Sci. Instrum., 86 (2015) 054902.
[13] T. Chen, S. Song, Y. Shen, K. Zhang, P. Jiang, Simultaneous Measurement of Thermal Conductivity and Heat Capacity Across Diverse Materials Using the Square-Pulsed Source (SPS) Technique, (2024) DOI: 10.48550/arXiv.2405.20870
[14] P. Jiang, X. Qian, R. Yang, A new elliptical-beam method based on time-domain

thermoreflectance (TDTR) to measure the in-plane anisotropic thermal conductivity and its comparison with the beam-offset method, Rev. Sci. Instrum., 89 (2018) 094902.

[15] P. Jiang, X. Qian, R. Yang, Tutorial: Time-domain thermoreflectance (TDTR) for thermal property characterization of bulk and thin film materials, J. Appl. Phys., 124 (2018) 161103.

[16] J. Yang, E. Ziade, A.J. Schmidt, Uncertainty analysis of thermoreflectance measurements, Rev. Sci. Instrum., 87 (2016) 014901.

[17] E.H. Ratcliffe, Thermal conductivities of fused and crystalline quartz, Br. J. Appl. Phys., 10 (1959) 22.

[18] D. Wang, H. Ban, P. Jiang, Spatially resolved lock-in micro-thermography (SR-LIT): A tensor analysis-enhanced method for anisotropic thermal characterization, Appl. Phys. Rev., 11 (2024).

[19] A.F. Birch, H. Clark, The thermal conductivity of rocks and its dependence upon temperature and composition, Am. J. Sci., 238 (1940) 529-558.

[20] G.W.C. Kaye, W.F. Higgins, J.E. Petavel, The thermal conductivity of vitreous silica, with a note on crystalline quartz, Proceedings of the Royal Society of London. Series A, Containing Papers of a Mathematical and Physical Character, 113 (1926) 335-351.

[21] Y. Touloukian, R. Powell, C. Ho, P. Klemens, Thermal Conductivity Nonmetallic Solids in The TPRC Data Series, 2, IFI/Plenum, New York, Washington, 1970.

[22] C.T. Anderson, The Heat Capacities of Quartz, Cristobalite and Tridymite at Low Temperatures1, J. Am. Chem. Soc., 58 (1936) 568-570.

# Supplementary Information

Comprehensive Measurement of Three-Dimensional Thermal Conductivity Tensor Using a Beam-Offset Square-Pulsed Source (BO-SPS) Approach

Tao Chen[1], Shangzhi Song[1], Puqing Jiang[1],*

[1]School of Power and Energy Engineering, Huazhong University of Science and Technology, Wuhan, Hubei 430074, China

**Section S1. Thermal Model Derivation**

*S1.1 Heat diffusion in a multilayered system with anisotropic thermal conductivities*

In this thermal model, we consider a general case of a multilayer system, where each layer has homogeneous but anisotropic thermal conductivities. The governing equation of heat diffusion is given by:

$$C\frac{\partial T}{\partial t} = k_{xx}\frac{\partial^2 T}{\partial x^2} + k_{yy}\frac{\partial^2 T}{\partial y^2} + k_{zz}\frac{\partial^2 T}{\partial z^2} + 2k_{xy}\frac{\partial^2 T}{\partial x \partial y} + 2k_{xz}\frac{\partial^2 T}{\partial x \partial z} + 2k_{yz}\frac{\partial^2 T}{\partial y \partial z} \quad (S1)$$

This parabolic partial differential equation can be simplified by applying Fourier transforms to the in-plane coordinates and time, $T(x,y,z,t) \leftrightarrow \Theta(u,v,z,\omega)$. Equation (S1) transforms into:

$$(iC\omega)\Theta = -4\pi^2(k_{xx}u^2 + 2k_{xy}uv + k_{yy}v^2)\Theta + 2i2\pi(k_{xz}u + k_{yz}v)\frac{\partial \Theta}{\partial z} + k_{zz}\frac{\partial^2 \Theta}{\partial z^2} \quad (S2)$$

where, $\Theta = \int_{-\infty}^{\infty}\int_{-\infty}^{\infty}\int_{-\infty}^{\infty} T e^{-i2\pi ux} dx\, e^{-i2\pi v}\, dy\, e^{-i\omega t} dt$

Equation (S2) can also be written more compactly as:

$$\frac{\partial^2 \Theta}{\partial z^2} + \lambda_2 \frac{\partial \Theta}{\partial z} - \lambda_1 \Theta = 0 \quad (S3)$$

where,

$$\lambda_1 = \frac{iC\omega}{k_{zz}} + \frac{4\pi^2(k_{xx}u^2 + 2k_{xy}uv + k_{yy}v^2)}{k_{zz}} \quad (S4)$$

$$\lambda_2 = 2i2\pi \frac{(k_{xz}u + k_{yz}v)}{k_{zz}} \quad (S5)$$

The general solution of Equation (S3) is

$$\Theta = e^{u^+ z}B^+ + e^{u^- z}B^- \quad (S6)$$

where $u^+$ and $u^-$ are the roots of the equation $x^2 + \lambda_2 x - \lambda_1 = 0$:

$$u^{\pm} = \frac{-\lambda_2 \pm \sqrt{(\lambda_2)^2 + 4\lambda_1}}{2} \tag{S7}$$

and $B^+, B^-$ are complex numbers to be determined.

The heat flux can be obtained from the temperature Equation (S6) and Fourier's law of heat conduction $Q = -k_{zz}\left(\frac{d\Theta}{dz}\right)$:

$$Q = -k_{zz} u^+ e^{u^+ z} B^+ - k_{zz} u^- e^{u^- z} B^- \tag{S8}$$

It is convenient to write Equations. (S6) and (S8) in matrix form:

$$\begin{bmatrix} \Theta \\ Q \end{bmatrix}_{n,z} = [\mathbf{N}]_n \begin{bmatrix} B^+ \\ B^- \end{bmatrix}_n \tag{S9}$$

where,

$$[\mathbf{N}]_n = \begin{bmatrix} 1 & 1 \\ -k_{zz} u^+ & -k_{zz} u^- \end{bmatrix}_n \begin{bmatrix} e^{u^+ z} & 0 \\ 0 & e^{u^- z} \end{bmatrix}_n \tag{S10}$$

Here, $n$ stands for the $n$-th layer of the multilayer system, and $z$ is the distance from the surface of the $n$-th layer.

The constants $B^+$ and $B^-$ for the $n$-th layer can also be obtained from the surface temperature and heat flux of that layer by setting $z = 0$ in Equation (S10) and performing its matrix inversion:

$$\begin{bmatrix} B^+ \\ B^- \end{bmatrix}_n = [\mathbf{M}]_n \begin{bmatrix} \Theta \\ Q \end{bmatrix}_{n,z=0} \tag{S11}$$

where,

$$[\mathbf{M}]_n = \frac{1}{k_{zz}(u^+ - u^-)} \begin{bmatrix} -k_{zz} u^- & -1 \\ k_{zz} u^+ & 1 \end{bmatrix} \tag{S12}$$

For heat flow across an interface, an interface conductance $G$ is defined. Therefore, the heat flux across an interface can be written as:

$$Q_{n,z=L} = Q_{n+1,z=0} = G(\Theta_{n,z=L} - \Theta_{n+1,z=0}) \tag{S13}$$

From Equation (S13), we also have:

$$\Theta_{n+1,z=0} = \Theta_{n,z=L} - \frac{1}{G} Q_{n,z=L} \tag{S14}$$

It is convenient to write Equations (S13) and (S14) in matrix form:

$$\begin{bmatrix} \Theta \\ Q \end{bmatrix}_{n+1,z=0} = [\mathbf{R}]_n \begin{bmatrix} \Theta \\ Q \end{bmatrix}_{n,z=L} \tag{S15}$$

where,

$$[\mathbf{R}]_n = \begin{bmatrix} 1 & -1/G \\ 0 & 1 \end{bmatrix}_n \tag{S16}$$

Here, $G_n$ represents the interfacial thermal conductivity between the $n$-th layer and the ($n$+1)-th layer.

The surface temperature and heat flux can thus be related to those at the bottom of the substrate as

$$\begin{bmatrix} \Theta \\ Q \end{bmatrix}_{n,z=L_n} = [N]_n [M]_n \cdots [R]_1 [N]_1 [M]_1 \begin{bmatrix} \Theta \\ Q \end{bmatrix}_{1,z=0} = \begin{bmatrix} A & B \\ C & D \end{bmatrix} \begin{bmatrix} \Theta \\ Q \end{bmatrix}_{1,z=0} \quad (S17)$$

In the model, the boundary condition at the bottom of the last layer is adiabatic, which means $C\Theta_{1,z=0} + DQ_{1,z=0} = 0$. The Green's function $\hat{G}$, which is essentially the detected temperature response due to the applied heat flux of unit strength, can thus be solved as

$$\hat{G}(u,v,\omega) = \frac{\Theta_{1,z=0}}{Q_{1,z=0}} = -\frac{D}{C} \quad (S18)$$

With the Green's function $\hat{G}$ determined, the detected temperature response is simply the product of $\hat{G}$ and the heat source function in the frequency domain.

*S1.2 Modeling of signals acquired in the experiments*

In the case of Gaussian profile laser heating modulated by a square wave, the surface heat flux is given by:

$$p_0(x,y,t) = \frac{2A_0}{\pi \sigma_{x_0} \sigma_{y_0}} e^{-\left(\frac{2x^2}{\sigma_{x_0}^2} + \frac{2y^2}{\sigma_{y_0}^2}\right)} \left(\frac{1}{2} + \frac{2}{\pi}\sum_{n=1}^{\infty} \frac{\sin(2\pi(2n-1)f_0 t)}{2n-1}\right) \quad (S19)$$

where, $A_0$ is the average power of the pump beam; $\sigma_{x_0}$ and $\sigma_{y_0}$ are the $1/e^2$ radii of the pump spot in the $x$ and $y$ directions, respectively; $f_0$ is the modulation frequency.

The Fourier transform of $p_0(x,y,t)$ over space and time is

$$P_0(u,v,\omega) = A_0 e^{-\frac{\pi^2 u^2 \sigma_{x_0}^2}{2}} e^{-\frac{\pi^2 v^2 \sigma_{y_0}^2}{2}} \left(\frac{\delta(\omega)}{2} + \frac{1}{\pi}\sum_{n=1}^{\infty} i\frac{(\delta(\omega+2\pi(2n-1)f_0) - \delta(\omega-2\pi(2n-1)f_0))}{2n-1}\right) \quad (S20)$$

where $\delta(x)$ is the Dirac delta function.

The detected temperature response is the product of the surface heat flux $P_0(u,v,\omega)$ and the Green's function $\hat{G}(u,v,\omega)$ in the frequency domain. The inverse Fourier transform yields the real-space distribution as

$$\theta(x,y,\omega) = \int_{-\infty}^{\infty} \int_{-\infty}^{\infty} P_0(u,v,\omega) \hat{G}(u,v,\omega) e^{i2\pi(ux+vy)} du dv \quad (S21)$$

Another continuous-wave laser beam with an offset distance $(x_c, y_c)$ from the pump beam is used to detect a weighted average of the transient temperature change as:

$$\Delta\theta(\omega) = \frac{2}{\pi \sigma_{x_1} \sigma_{y_1}} \int_{-\infty}^{\infty} \int_{-\infty}^{\infty} \theta(x,y,\omega) e^{-\frac{2(x-x_c)^2}{\sigma_{x_1}^2}} e^{-\frac{2(y-y_c)^2}{\sigma_{y_1}^2}} dx dy \quad (S22)$$

The integral of $\theta$ over $x$ and $y$ in Equation (S22) is the inverse Fourier transform of the probe beam with offsets, leaving an integral over $u$ and $v$ as

$$\Delta\theta(\omega) = \int_{-\infty}^{\infty} \int_{-\infty}^{\infty} A_0 \left(\frac{\delta(\omega)}{2} + \frac{1}{\pi}\sum_{n=1}^{\infty} i\frac{(\delta(\omega+2\pi(2n-1)f_0) - \delta(\omega-2\pi(2n-1)f_0))}{2n-1}\right) \hat{G}(u,v,\omega) e^{-\pi^2 u^2 w_x^2} e^{-\pi^2 v^2 w_y^2} e^{i2\pi(ux_c+vy_c)} du dv \quad (S23)$$

where: $w_x^2 = \frac{(\sigma_{x_0}^2 + \sigma_{x_1}^2)}{2}$, $w_y^2 = \frac{(\sigma_{y_0}^2 + \sigma_{y_1}^2)}{2}$.

The detected signal is the inverse Fourier transform of Equation (S23) as:

$$\Delta T(t) = A_0 \int_{-\infty}^{\infty} \int_{-\infty}^{\infty} \hat{G}(u,v,\omega) e^{-\pi^2 u^2 w_x^2} e^{-\pi^2 v^2 w_y^2} e^{i2\pi(ux_c + vy_c)} du dv \int_{-\infty}^{\infty} \left( \frac{\delta(\omega)}{2} + \frac{1}{\pi} \sum_{n=1}^{\infty} i \frac{(\delta(\omega + 2\pi(2n-1)f_0) - \delta(\omega - 2\pi(2n-1)f_0))}{2n-1} \right) e^{i\omega t} d\omega \quad \text{(S24)}$$

Equation (S24) can be further simplified to:

$$\Delta T(t) = \frac{A_0}{2} \int_{-\infty}^{\infty} \int_{-\infty}^{\infty} \hat{G}(u,v,0) e^{-\pi^2 u^2 w_x^2} e^{-\pi^2 v^2 w_y^2} e^{i2\pi(ux_c + vy_c)} du dv - 2A_0 \cdot \text{Re} \left\{ \sum_{n=1}^{\infty} \int_{-\infty}^{\infty} \int_{-\infty}^{\infty} \hat{G}(u,v, 2\pi(2n-1)f_0) \exp(-\pi^2 u^2 w_x^2) \exp(-\pi^2 v^2 w_y^2) e^{i2\pi(ux_c + vy_c)} du dv \frac{i e^{i2\pi(2n-1)f_0 t}}{\pi(2n-1)} \right\} \quad \text{(S25)}$$

where $\text{Re}\{z\}$ represents the real part of the complex number $z$.

**Section S2. Uncertainty Analysis of Multi-Parameter Extraction from Multi-Signal Fitting**

In processing the data, multiple parameters are extracted by simultaneously fitting different sets of experimental signals using the least-squares regression method. Mathematically, this involves minimizing the product of the root mean squared (RMS) differences between each set of experimental signals and their corresponding model predictions:

$$J = \prod_{j=1}^{M} \sqrt{\frac{1}{N_j} \sum_{i=1}^{N_j} \left(\frac{g_j(X_U, X_P, t_i)}{y_j(t_i)} - 1\right)^2} = \prod_{j=1}^{M} \text{RMS}_j \tag{S26}$$

At the best fit, the gradient of $J$ should be zero for every element in $X_U$:

$$\sum_{j=1}^{M} \left(\frac{\prod_{k \neq j} \text{RMS}_k}{2\text{RMS}_j}\right) \left(\frac{1}{N_j} \sum_{i=1}^{N_j} \frac{2(g_j(\hat{X}_U, \hat{X}_P, t_i) - y_j(t_i))}{y_j^2(t_i)} \frac{\partial g_j(\hat{X}_U, \hat{X}_P, t_i)}{\partial u_l}\right) = 0,$$

$$\text{for } l = 1, 2, \ldots, \text{length}(X_U) \tag{S27}$$

Here, $\hat{X}_P$ is a random group of the possible control parameters since these input parameters have uncertainties, and $\hat{X}_U$ is the corresponding group of fitting parameters that make the best fit. The uncertainties of the unknown parameters can be revealed from the distribution of all possible $\hat{X}_U$. Let us denote the mean values of all possible $\hat{X}_U$ and $\hat{X}_P$ as $X_U^0$ and $X_P^0$, respectively. The function $g_j(\hat{X}_U, \hat{X}_P, t_i)$ can be approximated by a first-order Taylor expansion around the point $(X_U^0, X_P^0)$ as:

$$g_j(\hat{X}_U, \hat{X}_P, t_i) \approx g_j(X_U^0, X_P^0, t_i) + \sum_{l=1}^{\text{length}(X_U)} \frac{g_j(\hat{X}_U, \hat{X}_P, t_i)}{\partial u_l}\bigg|_{X_U^0, X_P^0} (u_l^* - u_l^0)$$

$$+ \sum_{k=1}^{\text{length}(X_P)} \frac{\partial g_j(\hat{X}_U, \hat{X}_P, t_i)}{\partial p_k}\bigg|_{X_U^0, X_P^0} (p_k^* - p_k^0), \text{ for } j = 1, 2, \ldots, M \tag{S28}$$

Substituting Equation (S28) into Equation (S27) and neglecting the higher-order terms, we get:

$$\sum_{j=1}^{M} \left(\frac{\prod_{k \neq j} \text{RMS}_k}{\text{RMS}_j}\right) \left(\frac{1}{N_j} \sum_{i=1}^{N_j} \frac{1}{y_j^2(t_i)} \left(g_j(X_U^0, X_P^0, t_i) - y_j(t_i) + \sum_{l=1}^{\text{length}(X_U)} \frac{g_j(\hat{X}_U, \hat{X}_P, t_i)}{\partial u_l}\bigg|_{X_U^0, X_P^0} (u_l^* - u_l^0)\right.\right.$$

$$\left.\left.+ \sum_{k=1}^{\text{length}(X_P)} \frac{\partial g_j(\hat{X}_U, \hat{X}_P, t_i)}{\partial p_k}\bigg|_{X_U^0, X_P^0} (p_k^* - p_k^0)\right) \left(\frac{\partial g_j(\hat{X}_U, \hat{X}_P, t_i)}{\partial u_l}\bigg|_{X_U^0, X_P^0}\right)\right) = 0,$$

$$\text{for } l = 1, 2, \ldots, \text{length}(X_U) \tag{S29}$$

Eq. (S29) could be re-written in a matrix format as:

$$\sum_{j=1}^{M} \left(\frac{\prod_{k \neq j} \text{RMS}_k}{N_j \text{RMS}_j}\right) J_{U,j}^T G_j [F_j - E_j + J_{U,j}(\hat{X}_U - X_U^0) + J_{P,j}(\hat{X}_P - X_P^0)] = 0 \tag{S30}$$

where $G_j$ is the diagonal matrix $G_j = diag(\frac{1}{y_j^2(t_1)}, \frac{1}{y_j^2(t_2)}, \cdots, \frac{1}{y_j^2(t_{N_j})})$, $E_j$ is the column vector of the $j$-th set of measured signals, and $F_j$ is the corresponding column vector of the signals evaluated by the thermal model at $(X_U^0, X_P^0)$. $J_{U,j}$ and $J_{P,j}$ are the Jacobian matrices of the function $F_j$ for variables $X_U$ and $X_P$, respectively:

$$J_{U,j} = \begin{pmatrix} \frac{\partial g_j(\hat{X}_U, \hat{X}_P, t_1)}{\partial u_1}|_{X_U^0, X_P^0} & \cdots & \frac{\partial g_j(\hat{X}_U, \hat{X}_P, t_1)}{\partial u_{\text{length}(X_U)}}|_{X_U^0, X_P^0} \\ \vdots & \ddots & \vdots \\ \frac{\partial g_j(\hat{X}_U, \hat{X}_P, t_{N_j})}{\partial u_1}|_{X_U^0, X_P^0} & \cdots & \frac{\partial g_j(\hat{X}_U, \hat{X}_P, t_{N_j})}{\partial u_{\text{length}(X_U)}}|_{X_U^0, X_P^0} \end{pmatrix} \quad (S31)$$

and

$$J_{P,j} = \begin{pmatrix} \frac{\partial g_j(\hat{X}_U, \hat{X}_P, t_1)}{\partial p_1}|_{X_U^0, X_P^0} & \cdots & \frac{\partial g_j(\hat{X}_U, \hat{X}_P, t_1)}{\partial p_{\text{length}(X_P)}}|_{X_U^0, X_P^0} \\ \vdots & \ddots & \vdots \\ \frac{\partial g_j(\hat{X}_U, \hat{X}_P, t_{N_j})}{\partial p_1}|_{X_U^0, X_P^0} & \cdots & \frac{\partial g_j(\hat{X}_U, \hat{X}_P, t_{N_j})}{\partial p_{\text{length}(X_P)}}|_{X_U^0, X_P^0} \end{pmatrix} \quad (S32)$$

Equation (S30) can be rearranged as:

$$\sum_{j=1}^{M} \left( \frac{\prod_{k \neq j} \text{RMS}_k}{N_j \text{RMS}_j} \right) J_{U,j}^T G_j (E_j - F_j) - \sum_{j=1}^{M} \left( \frac{\prod_{k \neq j} \text{RMS}_k}{N_j \text{RMS}_j} \right) J_{U,j}^T G_j J_{P,j} (\hat{X}_P - X_P^0)$$

$$= \sum_{j=1}^{M} \left( \frac{\prod_{k \neq j} \text{RMS}_k}{N_j \text{RMS}_j} \right) J_{U,j}^T G_j J_{U,j} (\hat{X}_U - X_U^0) \quad (S33)$$

Let us denote

$$\Sigma_{UU} = \sum_{j=1}^{M} \left( \frac{\prod_{k \neq j} \text{RMS}_k}{N_j \text{RMS}_j} \right) J_{U,j}^T G_j J_{U,j} \quad (S33a)$$

$$\Sigma_{UP} = \sum_{j=1}^{M} \left( \frac{\prod_{k \neq j} \text{RMS}_k}{N_j \text{RMS}_j} \right) J_{U,j}^T G_j J_{P,j} \quad (S33b)$$

When $\Sigma_{UU}$ is non-singular, we can explicitly express $\hat{X}_U$ as:

$$\hat{X}_U = \Sigma_{UU}^{-1} \sum_{j=1}^{M} \left( \frac{\prod_{k \neq j} \text{RMS}_k}{N_j \text{RMS}_j} \right) J_{U,j}^T G_j (E_j - F_j) - \Sigma_{UU}^{-1} \Sigma_{UP} (\hat{X}_P - X_P^0) + X_U^0 \quad (S34)$$

The distributions of elements in $\hat{X}_U$ can be obtained by calculating its covariance matrix. Since $E_j$ and $\hat{X}_P$ are independent vectors, the covariance matrix of $\hat{X}_U$ can be expressed as

$$\text{Var}[\hat{X}_U] = \Sigma_{UU}^{-1} \left[ \sum_{j=1}^{M} \left( \frac{\prod_{k \neq j} \text{RMS}_k}{N_j \text{RMS}_j} \right)^2 J_{U,j}^T G_j \text{Var}[E_j - F_j] G_j^T J_{U,j} \right] \Sigma_{UU}^{-1}$$

$$+ \Sigma_{UU}^{-1} \Sigma_{UP} \text{Var}[\hat{X}_P] \Sigma_{UP}^T \Sigma_{UU}^{-1} \quad (S35)$$

Here, $\text{Var}[E_j]$ is an $N_j$–by–$N_j$ diagonal matrix with the $i$-th component being $\left( y_j(t_i) - g_j(X_U^0, X_P^0, t_i) \right)^2$, and $\text{Var}[\hat{X}_P]$ is a length($X_P$)-by-length($X_P$) diagonal matrix with the $k$-th component being $\sigma_{p_k}^2$.

Equation (S35) is the error propagation formula, which is a summation of two terms: the first term is the uncertainty from the experimental noise and fitting quality, and the second term is the uncertainty propagated from the errors of the control variables. The covariance matrix $\text{Var}[\hat{X}_U]$ takes the format:

$$\text{Var}[\hat{X}_U] = \begin{bmatrix} \sigma_{u_1}^2 & \text{cov}[u_1, u_2] & \cdots \\ \text{cov}[u_2, u_1] & \sigma_{u_2}^2 & \cdots \\ \vdots & \vdots & \ddots \end{bmatrix} \tag{S36}$$

where the elements on the principal diagonal $\sigma_{u_1}, \sigma_{u_2}, \ldots, \sigma_{u_{\text{length}(X_U)}}$ are the variances of the unknown parameters $u_1, u_2, \ldots, u_{\text{length}(X_U)}$; the off-diagonal ones $\text{cov}[u_i, u_j]$ are the covariances of $u_i$ and $u_j$. If $\text{cov}[u_i, u_j] = 0$, this means the variables $u_i$ and $u_j$ are entirely independent of each other.